\documentclass[twoside,twocolumn,9pt]{article}
\usepackage{extsizes}
\usepackage[super,sort&compress,comma,numbers]{natbib} 
\usepackage[version=3]{mhchem}
\usepackage[left=1.5cm, right=1.5cm, top=1.785cm, bottom=2.0cm]{geometry}
\usepackage{balance}
\usepackage{times, mathptmx}
\usepackage{sectsty}
\usepackage{graphicx}
\usepackage{lastpage}
\usepackage[format=plain,justification=justified,singlelinecheck=false,font={stretch=1.125,small,sf},labelfont=bf,labelsep=space]{caption}
\usepackage{float}
\usepackage{fancyhdr}
\usepackage{fnpos}
\usepackage[english]{babel}
\addto{\captionsenglish}{%
  
}
\usepackage{array}
\usepackage{droidsans}
\usepackage{charter}
\usepackage[T1]{fontenc}
\usepackage[usenames,dvipsnames,table]{xcolor}
\usepackage{setspace}
\usepackage[compact]{titlesec}
\usepackage{hyperref}
\usepackage{tabularx}
\usepackage{dblfloatfix}
\usepackage{transparent}
\usepackage[usetitle]{rsc}

\usepackage{epstopdf}
\usepackage{blindtext}

\definecolor{cream}{RGB}{222,217,201}

\begin{document}

\pagestyle{fancy}
\thispagestyle{plain}
\fancypagestyle{plain}{
\renewcommand{\headrulewidth}{0pt}
}

\makeFNbottom
\makeatletter
\renewcommand\LARGE{\@setfontsize\LARGE{15pt}{17}}
\renewcommand\Large{\@setfontsize\Large{12pt}{14}}
\renewcommand\large{\@setfontsize\large{10pt}{12}}
\renewcommand\footnotesize{\@setfontsize\footnotesize{7pt}{10}}
\makeatother

\renewcommand{\thefootnote}{\fnsymbol{footnote}}
\renewcommand\footnoterule{\vspace*{1pt}%
\color{cream}\hrule width 3.5in height 0.4pt \color{black}\vspace*{5pt}} 
\setcounter{secnumdepth}{5}

\makeatletter 
\renewcommand\@biblabel[1]{#1}            
\renewcommand\@makefntext[1]%
{\noindent\makebox[0pt][r]{\@thefnmark\,}#1}
\makeatother 
\renewcommand{\figurename}{\small{Fig.}~}
\sectionfont{\sffamily\Large}
\subsectionfont{\normalsize}
\subsubsectionfont{\bf}
\setstretch{1.125} 
\setlength{\skip\footins}{0.8cm}
\setlength{\footnotesep}{0.25cm}
\setlength{\jot}{10pt}
\titlespacing*{\section}{0pt}{4pt}{4pt}
\titlespacing*{\subsection}{0pt}{15pt}{1pt}

\fancyfoot{}
\fancyfoot[LO,RE]{\vspace{-7.1pt}\includegraphics[height=9pt]{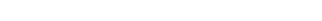}}
\fancyfoot[CO]{\vspace{-7.1pt}\hspace{13.2cm}\includegraphics{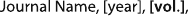}}
\fancyfoot[CE]{\vspace{-7.2pt}\hspace{-14.2cm}\includegraphics{head_foot/RF}}
\fancyfoot[RO]{\footnotesize{\sffamily{1--\pageref{LastPage} ~\textbar  \hspace{2pt}\thepage}}}
\fancyfoot[LE]{\footnotesize{\sffamily{\thepage~\textbar\hspace{3.45cm} 1--\pageref{LastPage}}}}
\fancyhead{}
\renewcommand{\headrulewidth}{0pt} 
\renewcommand{\footrulewidth}{0pt}
\setlength{\arrayrulewidth}{1pt}
\setlength{\columnsep}{6.5mm}
\setlength\bibsep{1pt}

\makeatletter 
\newlength{\figrulesep} 
\setlength{\figrulesep}{0.5\textfloatsep} 

\newcommand{\topfigrule}{\vspace*{-1pt}%
\noindent{\color{cream}\rule[-\figrulesep]{\columnwidth}{1.5pt}} }

\newcommand{\botfigrule}{\vspace*{-2pt}%
\noindent{\color{cream}\rule[\figrulesep]{\columnwidth}{1.5pt}} }

\newcommand{\dblfigrule}{\vspace*{-1pt}%
\noindent{\color{cream}\rule[-\figrulesep]{\textwidth}{1.5pt}} }

\makeatother

\twocolumn[
  \begin{@twocolumnfalse}
{\includegraphics[height=30pt]{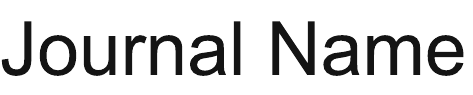}\hfill\raisebox{0pt}[0pt][0pt]{\includegraphics[height=55pt]{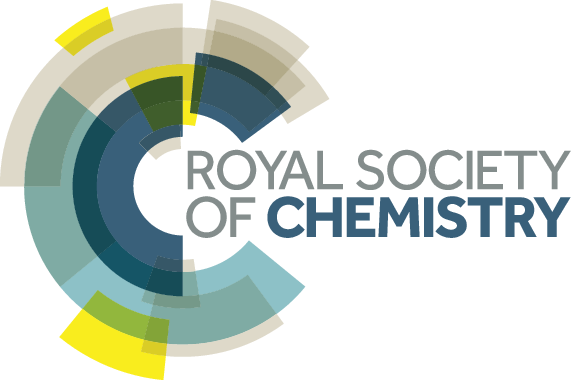}}\\[1ex]
\includegraphics[width=18.5cm]{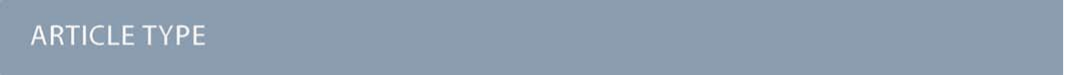}}\par
\vspace{1em}
\sffamily
\begin{tabular}{m{4.5cm} p{13.5cm} }

\includegraphics{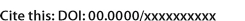} & \noindent\LARGE{\textbf{Monolithic Selenium/Silicon Tandem Solar Cells}} \\
\vspace{0.3cm} & \vspace{0.3cm} \\

 & \noindent\large{Rasmus Nielsen,$^{\ast}$\textit{$^{a}$} 
Andrea Crovetto,\textit{$^{b}$} Alireza Assar,\textit{$^{b}$}, Ole Hansen,\textit{$^{b}$} Ib Chorkendorff,\textit{$^{a}$} and Peter C. K. Vesborg\textit{$^{a}$}} \\

\includegraphics{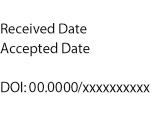} & \noindent\normalsize{Selenium is experiencing renewed interest as a promising candidate for the wide bandgap photoabsorber in tandem solar cells. However, despite the potential of selenium-based tandems to surpass the theoretical efficiency limit of single junction devices, such a device has never been demonstrated. In this study, we present the first monolithically integrated selenium/silicon tandem solar cell. Guided by device simulations, we investigate various carrier-selective contact materials and achieve encouraging results, including an open-circuit voltage of V$_\text{oc}$=1.68 V from suns-\textit{V}$_\text{oc}$ measurements. The high open-circuit voltage positions selenium/silicon tandem solar cells as serious contenders to the industrially dominant single junction technologies. Furthermore, we quantify a pseudo fill factor of more than 80\% using injection-level-dependent open-circuit voltage measurements, indicating that a significant fraction of the photovoltaic losses can be attributed to parasitic series resistance. This work provides valuable insights into the key challenges that need to be addressed for realizing higher efficiency selenium/silicon tandem solar cells.}\\

\end{tabular}

 \end{@twocolumnfalse} \vspace{0.6cm}

  ]

\renewcommand*\rmdefault{bch}\normalfont\upshape
\rmfamily
\section*{}
\vspace{-1cm}


\footnotetext{\textit{$^{a}$~SurfCat, DTU Physics, Technical University of Denmark, DK-2800 Kgs. Lyngby, Denmark.}}
\footnotetext{\textit{$^{b}$~DTU Nanolab, National Center for Nano Fabrication and Characterization, Technical University of Denmark, DK-2800, Kgs. Lyngby, Denmark.}}
\footnotetext{\textit{$^{\ast}$~ E-mail: raniel@dtu.dk}}

\footnotetext{\dag~Electronic Supplementary Information (ESI) available: Additional figures. See DOI: 00.0000/xxxxxxxxxx/}




\section*{Introduction}

Silicon-based photovoltaic (PV) technologies have dominated the commercial PV market for more than half a century, owing to their low cost and high reliability\cite{green2019a}. However, the power conversion efficiency (PCE) of silicon solar cells has now reached 26.8\%, which is approaching the theoretical upper limit of single junction devices\cite{green2023a, shockley1961a, richter2013a}. In order to further accelerate the deployment of PV, research must focus on high-efficiency concepts that effectively reduce the area-related balance-of-system costs\cite{creutzig2017a}. Among these concepts, tandem solar cells have emerged as the most promising technology for achieving efficiencies significantly beyond the practical limits of single junction solar cells\cite{martinho2021a}. Ideally, a tandem device combines bandgaps of \textit{E}$_\text{g, low}\approx0.9$ eV and \textit{E}$_\text{g, high}\approx1.6$ eV\cite{kirchartz2018a}, but silicon is arguably the most mature and technologically relevant candidate for the bottom cell with its bandgap of \textit{E}$_\text{g}=1.1$ eV. The wide bandgap partner material to integrate in tandem with silicon should then have a bandgap in the range of \textit{E}$_\text{g}\sim1.75$ eV, but only a limited number of candidates have been proposed\cite{al-ashouri2020a, mariotti2023a, yamaguchi2018a, hajijafarassar2020a, martinho2020a}, none of which have demonstrated both high performance, low cost, and long-term stability\cite{duan2023a, bobela2017a}.

\begin{figure*}[t!]
    \centering
    \includegraphics[width=\textwidth,trim={0 0 0 0},clip]{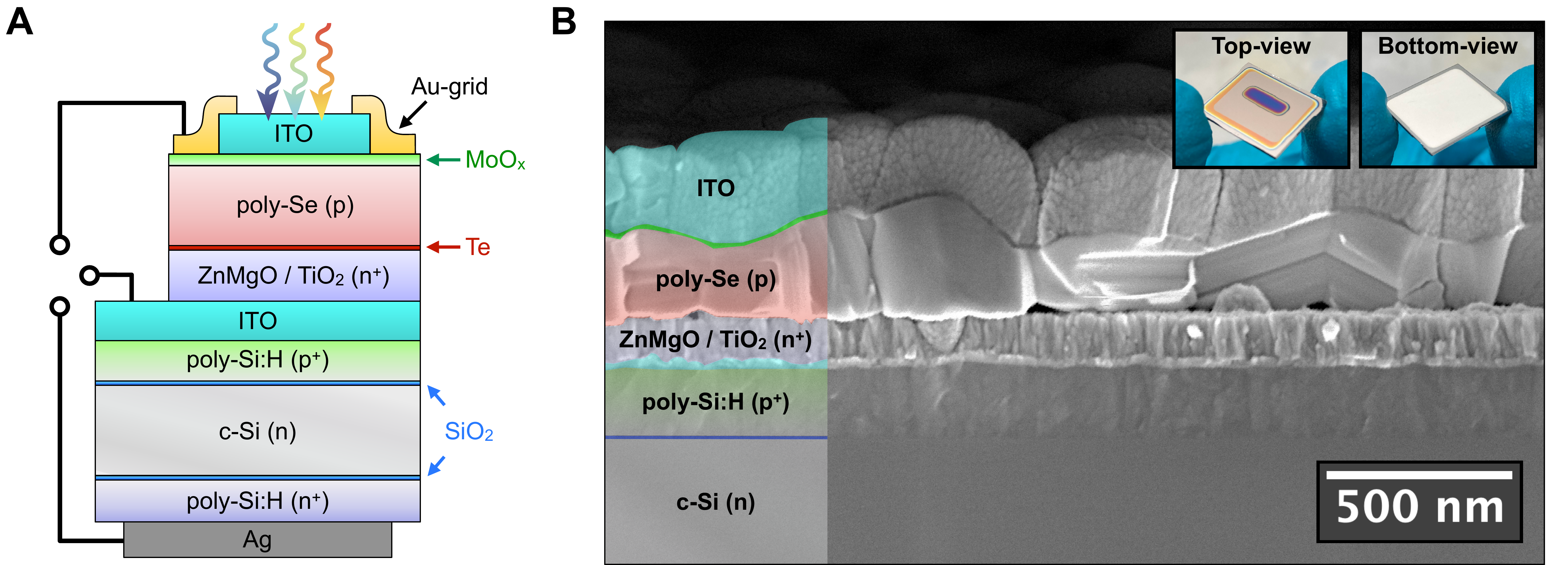}\newline
    \caption{Device architecture of the three-terminal monolithic selenium/silicon tandem solar cell. (\textbf{A}) Schematic illustration of the device structure using either ZnMgO or TiO$_\text{2}$ as the n-type contact. (\textbf{B}) Cross-sectional SEM-image of the tandem solar cell. Photographs of the device are also shown.}
    \label{fig:Figure1}
\end{figure*}

Selenium, the world's oldest PV material, is experiencing renewed interest as a promising wide bandgap photoabsorber for the top cell in tandem devices\cite{zhu2019a, yan2022a, liu2020b, fu2022a, lu2022a, nielsen2023a}. This inorganic semiconductor has a reported direct bandgap between 1.8 and 2.0 eV in its trigonal phase, which may be tuned within the range of 1.2-1.8 eV through alloying with tellurium\cite{hadar2019b, zheng2022a}. Furthermore, selenium features a very high absorption coefficient (>10$^\text{5}$ cm$^\text{-1}$) in the visible region\cite{nielsen2022a}, long-term air-stability\cite{liu2020a, zhu2016a}, and its single-element composition and low melting point of 220$^\circ$C makes processing simple, potentially low cost, and compatible with most bottom cells. Nevertheless, the research progress on selenium plateaued with an efficiency record of PCE$=$5.0\% achieved by Nakada et. al. in 1985\cite{nakada1985a}. This record remained unchanged for more than 30 years, until Todorov et al. introduced a redesigned device architecture in 2017, demonstrating a new record efficiency of PCE$=$6.5\%\cite{todorov2017a}. The next advancement involved replacing the non-transparent Au-electrode with a transparent conductive oxide (TCO) and thus realizing a bifacial device, which was accomplished in 2021 by Youngman et al. with a reported PCE$=$5.2\%\cite{youngman2021a}. However, to date, no reports have been published demonstrating an actual selenium/silicon tandem device.

In this work, we successfully fabricate the first monolithic selenium/silicon tandem solar cells. The initial tandem device uses ZnMgO as the electron-selective contact material, demonstrating an encouraging open-circuit voltage of V$_\text{oc}$=1.68 V. However, we observe that severe charge carrier transport losses limit the overall device performance. Using SCAPS-1D\cite{burgelman2000a}, we simulate the energy band diagrams of the heterostructure, and identify a critical transport barrier that is restricting the flow of electrons. To reduce the barrier height, we introduce TiO$_\text{2}$ as a replacement for ZnMgO, resulting in a 10-fold increase in the overall power conversion efficiency of the tandem to PCE=2.7\%. In parallel to the tandem devices, we fabricate and characterize bifacial single junction selenium solar cells to gain more comprehensive insights into the dependence of device polarity. Finally, we outline strategies for further improving the device performance to realize higher efficiency selenium/silicon tandem solar cells.\\

\section*{Experimental details}

The silicon bottom cells fabricated and used in this work comprise a double-sided tunnel oxide passivated contact (TOPCon) structure. First, a 1.4$\pm$0.1 nm SiO$_\text{2}$ tunnel oxide is grown on an n-type double-side polished silicon wafer using hot nitric acid oxidation. Immediately after the tunnel oxide has been grown, $\sim$200 nm polycrystalline silicon contacts are deposited using low pressure chemical vapor deposition (LPCVD). The poly-Si contacts are relatively thick as to protect the photoabsorber from sputter damage of the subsequently deposited thin-films. The dopant sources used during the LPCVD-processes are diborane for the p-type contact, and phosphine for the n-type contact. A graded doping profile is achieved by gradually increasing the flow of the dopant sources to minimize bulk interdiffusion through the tunnel oxide to the c-Si. The wafers are then annealed at 850$^\circ$C for 20 min in N$_\text{2}$ to promote dopant diffusion and activation. To further improve the passivation quality, a layer of $\sim$75 nm hydrogenated SiN$_\text{x}$:H is deposited using plasma-enhanced chemical vapor deposition (PECVD). After depositing the SiN$_\text{x}$:H, the devices are annealed in a tube furnace at 400$^\circ$C for 30 min to drive atomic hydrogen into the carrier-selective contacts, after which the SiN$_\text{x}$:H-layers are removed using buffered hydrofluoric acid. Then, a 25$\pm$4 nm indium tin oxide (ITO) recombination layer is deposited using RF magnetron sputtering with an elevated substrate temperature of 300$^\circ$C, and the wafers are finally diced to the dimensions 16 mm x 14 mm using a laser micromachining tool. The influence of the hydrogenation and sputter deposition steps on the photovoltaic performance has been quantified using photoconductance decay measurements presented in Fig. S.3 and S.4 along with batch statistics of the silicon bottom cells before dicing in Fig. S.5.

\begin{figure*}[t!]
    \centering
    \includegraphics[width=\textwidth,trim={0 0 0 0},clip]{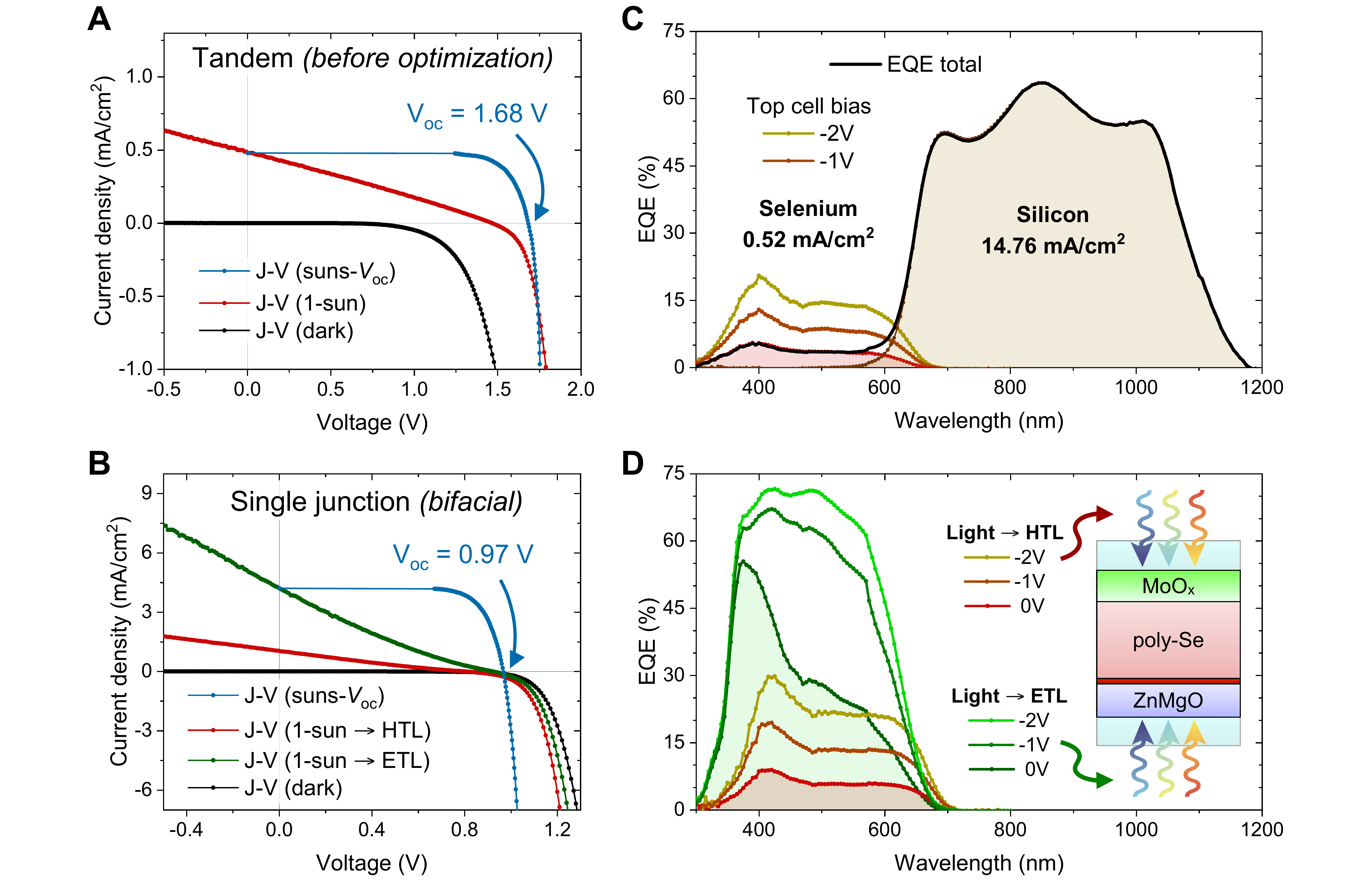} 
    \caption{Photovoltaic device performance of the ZnMgO-based selenium/silicon tandem and parallel-processed bifacial selenium single junction solar cells. (\textbf{A}) Current-voltage (\textit{J-V}) curves measured under 1-sun and dark conditions, accompanied by the \textit{J-V} curve reconstructed from suns-\textit{V}$_\text{oc}$ measurements. (\textbf{B}) Equivalent \textit{J-V} characterization for the bifacial single junction, with the device illuminated through either the hole- or electron transport layer (HTL or ETL). (\textbf{C}) External quantum efficiency (EQE) spectra of each subcell measured under short-circuit conditions, along with the spectra of the selenium top cell under reverse bias conditions. The AM1.5G-equivalent short-circuit current density of each subcell has been integrated and highlighted in bold. (\textbf{D}) EQE spectra of the bifacial single junction illuminated through either the HTL or ETL under short-circuit and reverse applied biases.}
    \label{fig:Figure2}
\end{figure*}

The selenium top cells are fabricated within a designated area of the diced bottom cells, intentionally leaving the edge of the ITO recombination layer exposed. First, a $\sim$65 nm layer of either ZnMgO or TiO$_\text{2}$ is sputter deposited as the electron-selective contact material. ZnMgO is RF sputter deposited from a ceramic Zn$_\text{0.8}$Mg$_\text{0.2}$O-target at room temperature in a reactive atmosphere of Ar/O$_\text{2}$ (60/0.6 sccm, $p$=5 mTorr) followed by an annealing step for 1 hr at 500$^\circ$C under high vacuum conditions. TiO$_\text{2}$ is DC sputter deposited from a metallic Ti-target in a reactive atmosphere of Ar/O$_\text{2}$ (60/6 sccm, $p$=5 mTorr) with an elevated substrate temperature of 400$^\circ$C. Once the samples have cooled down to room temperature, vacuum is broken and the samples are transferred to a custom-built thermal evaporator. Here, an ultrathin $\sim$1 nm layer of tellurium is deposited, immediately followed by the deposition of 300 nm of selenium, while keeping the samples at room temperature. Tellurium is used as an adhesion layer critical for improving the adhesion and uniformity of selenium during the crystallization process\cite{nakada1985a}. Selenium is crystallized by thermally annealing the samples in air at 190$^\circ$C for 4 min in a pre-heated aluminium oven. However, in the case of the ZnMgO-based devices on silicon, an additional pre-annealing step at 100$^\circ$C for 2 min was necessary to avoid the formation of large crater-like holes in the selenium thin-film, as shown in Fig. S.7. In contrast, this step was not required for TiO$_\text{2}$-based devices or single junctions fabricated on 1.1 mm thick glass, suggesting that the combined influence of the more abrupt heating through silicon and the different surface energy of ZnMgO may be responsible for the crater formation. Following the crystallization of the selenium thin-film, a 15 nm layer of MoO$_\text{x}$ (2<x<3, Ar/O$_\text{2}$=60/0.6 sccm, $p$=5 mTorr) and $\sim$250 nm of ITO (Ar/O$_\text{2}$=40/0.3 sccm, $p$=3 mTorr) are RF sputter deposited. Due to the low melting point of selenium, the substrate temperature is carefully maintained at 100$^\circ$C during the ITO deposition. As this is lower than the 300$^\circ$C used during the deposition of the ITO recombination layer on silicon, the resistivity of the ITO thin-film is expected to increase\cite{meng1998a, tuna2010a}. To compensate for the expected ohmic loss, an Au-grid was sputter deposited. In parallel to the fabrication of the tandem devices, bifacial single junction selenium solar cells were processed on ITO-coated glass substrates.\\

\section*{Results \& discussion}

The two selenium/silicon tandem structures fabricated in this work are schematically illustrated in Fig. \ref{fig:Figure1}A, using either ZnMgO or TiO$_\text{2}$ as the n-type contact in the selenium top cell. The accessible area of the ITO recombination layer is used as a third terminal, facilitating the characterization of the individual subcells. A cross-sectional scanning electron microscopy (SEM) image of the champion tandem device is shown in Fig. \ref{fig:Figure1}B.

\begin{figure*}[t!]
    \centering
    \includegraphics[width=\textwidth,trim={0 0 0 0},clip]{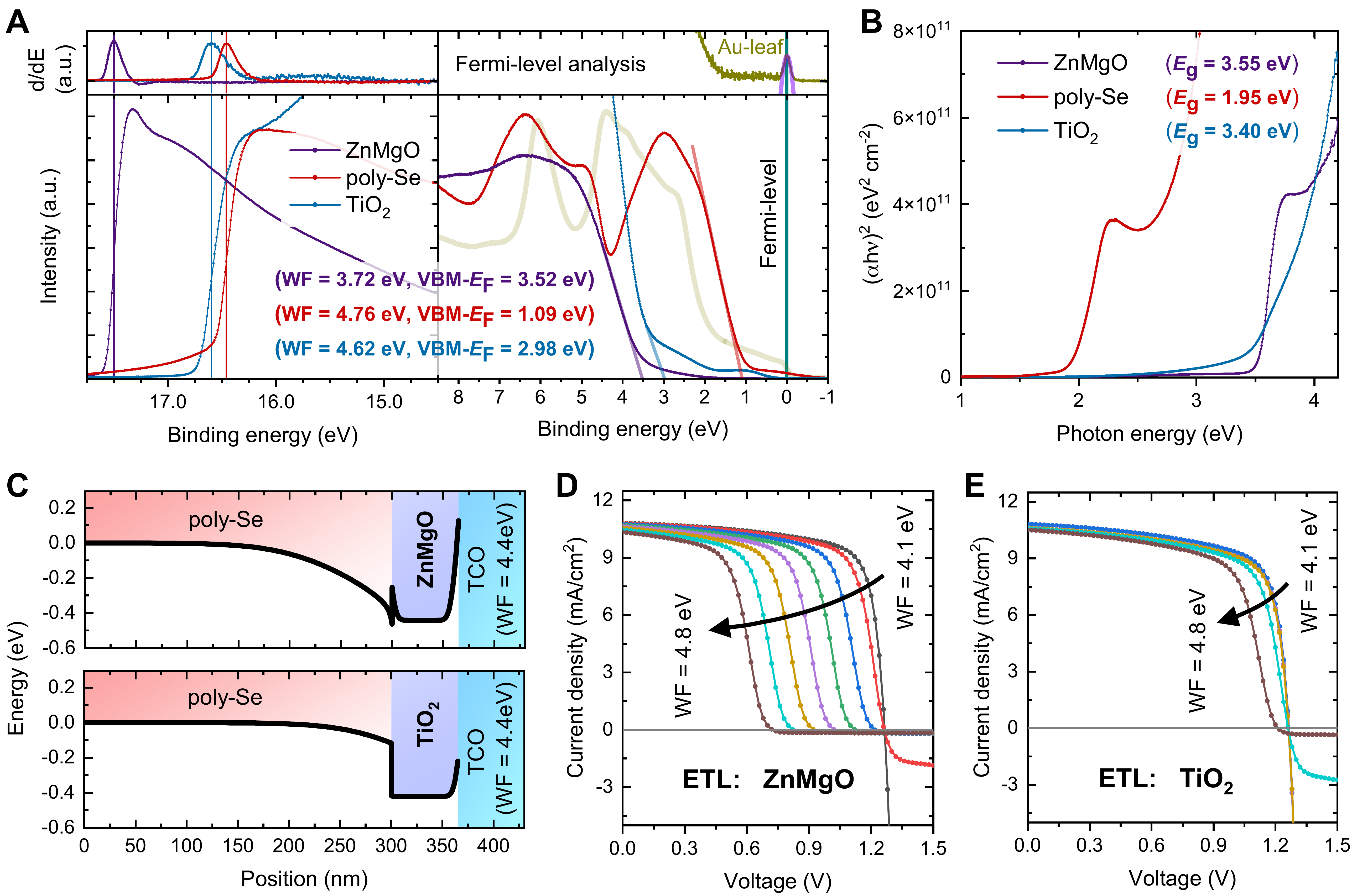}
    \caption{Characterization of energy band positions and simulated device performance. (\textbf{A}) Ultraviolet photoemission spectroscopy (UPS) measurements of ZnMgO, TiO$_\text{2}$ and poly-Se thin-films. The valence band offsets relative to the Fermi-level are extrapolated from the leading edges of the spectra, and the work functions are calculated from the secondary electron cutoffs. A small bias of -5V is applied to the samples to deconvolute the work function of the surface from that of the analyzer, and a gold leaf is mounted in electrical contact with each sample to accurately determine the Fermi-level. (\textbf{B}) Tauc plots derived from UV-vis transmission measurements. The optical bandgap of each thin-film is linearly extrapolated from the absorption onsets. (\textbf{C}) Simulated energy band diagrams of the poly-Se/ETL/TCO heterostructure under open-circuit conditions, featuring an arbitrary TCO with a work function of WF=4.4 eV. (\textbf{D}) Simulated \textit{J-V} curve sensitivity of the ZnMgO-based single junction device as the TCO work function is varied from 4.1 eV to 4.8 eV. (\textbf{E}) The equivalent sensitivity analysis of the TiO$_\text{2}$-based single junction device.}
    \label{fig:Figure3}
\end{figure*}

The current-voltage (\textit{J-V}) characteristics of the ZnMgO-based tandem and the bifacial selenium single junction are presented in Fig. \ref{fig:Figure2}A and B, respectively. Using suns-\textit{V}$_\text{oc}$, where no current is drawn from the cell, the measured open-circuit voltage is V$_\text{oc}=$1.68 V for the tandem and V$_\text{oc, Se}=$0.97 V for the parallel-processed top cell. Consequently, the silicon bottom cell must be delivering V$_\text{oc, Si}=$0.71 V, which perfectly matches the photovoltaic performance analysis of the bottom cells presented in Fig. S.5. However, both the tandem and the parallel-processed top cell suffer from significant carrier transport losses, as evident from the standard \textit{J-V} curves. In the case of the tandem device, these losses reduce the open-circuit voltage to V$_\text{oc}=$1.47 V and the fill-factor to FF$=$25\%. The consistent presence of these parasitic losses in both devices implies that the primary issue lies with the top cell. Another concern is the low short-circuit current density J$_\text{sc}<$1 mA/cm$^\text{2}$ measured when illuminating through the hole transport layer (HTL).

The bifaciality of the parallel-processed single junction allows for investigating the polarity dependence of the selenium top cell. By illuminating through the electron transport layer (ETL), more photons are absorbed in the vicinity of the carrier-separating p-n junction, leading to a significant improvement in the short-circuit current density by a factor of 4. This finding has been previously reported for bifacial selenium solar cells by Youngman et al., who attributed the result to low carrier diffusion lengths and lifetimes\cite{youngman2021a}. Moreover, as the current density increases, the \textit{J-V} curve becomes more S-shaped, suggesting the presence of a carrier transport barrier\cite{saive2019a}.

To better understand the carrier collection losses and polarity dependence, external quantum efficiency (EQE) spectra of the tandem and of the bifacial selenium single junction are shown in Fig. \ref{fig:Figure2}C and D, respectively. The integrated short-circuit current densities of the individual subcells show that the selenium top cell is limiting the overall current of the tandem device. By applying a reverse bias to the top cell using the accessible third terminal, the collection efficiency significantly improves. This indicates that under short-circuit conditions, the selenium thin-film is not fully depleted, and the photo-excited carriers generated beyond a diffusion length from the depletion region edge are lost to recombination. Similar observations are made for the parallel-processed single junction shown in Fig. \ref{fig:Figure2}D. However, by illuminating through the ETL and applying a reverse bias, a much higher collection efficiency comparable to state-of-the-art selenium single junction solar cells is achieved. These findings suggest that by inverting the device polarity and either depleting more of the selenium thin-film or significantly improving the diffusion length, short-circuit current densities on the order of $\sim$10 mA/cm$^\text{2}$ can be attained. To realize even higher current densities, the implementation of anti-reflecting coatings and other light trapping techniques should be considered.

The bifacial selenium solar cells presented in this work are fabricated in a process similar to that presented in our previously published work\cite{youngman2021a}. The only notable difference lies in the TCO on which ZnMgO is deposited. Specifically, the devices using fluorine doped tin oxide (FTO) instead of ITO show no S-shaped kink in the \textit{J-V} curve, which indicates that the carrier transport barrier could be attributed to the ZnMgO/TCO-interface. ZnMgO was first suggested for use in selenium thin-film solar cells by Todorov et al.\cite{todorov2017a}, and later played a key role in achieving the record open-circuit voltage of V$_\text{oc}=$0.99 V\cite{nielsen2022a}. However, all ZnMgO-based selenium solar cells reported in literature have been synthesized on FTO-coated glass, whereas the devices fabricated on ITO use TiO$_\text{2}$ instead of ZnMgO as the n-type contact material\cite{kunioka1982a, nakada1985a}.

\begin{figure*}[t!]
    \centering
    \includegraphics[width=\textwidth,trim={0 0 0 0},clip]{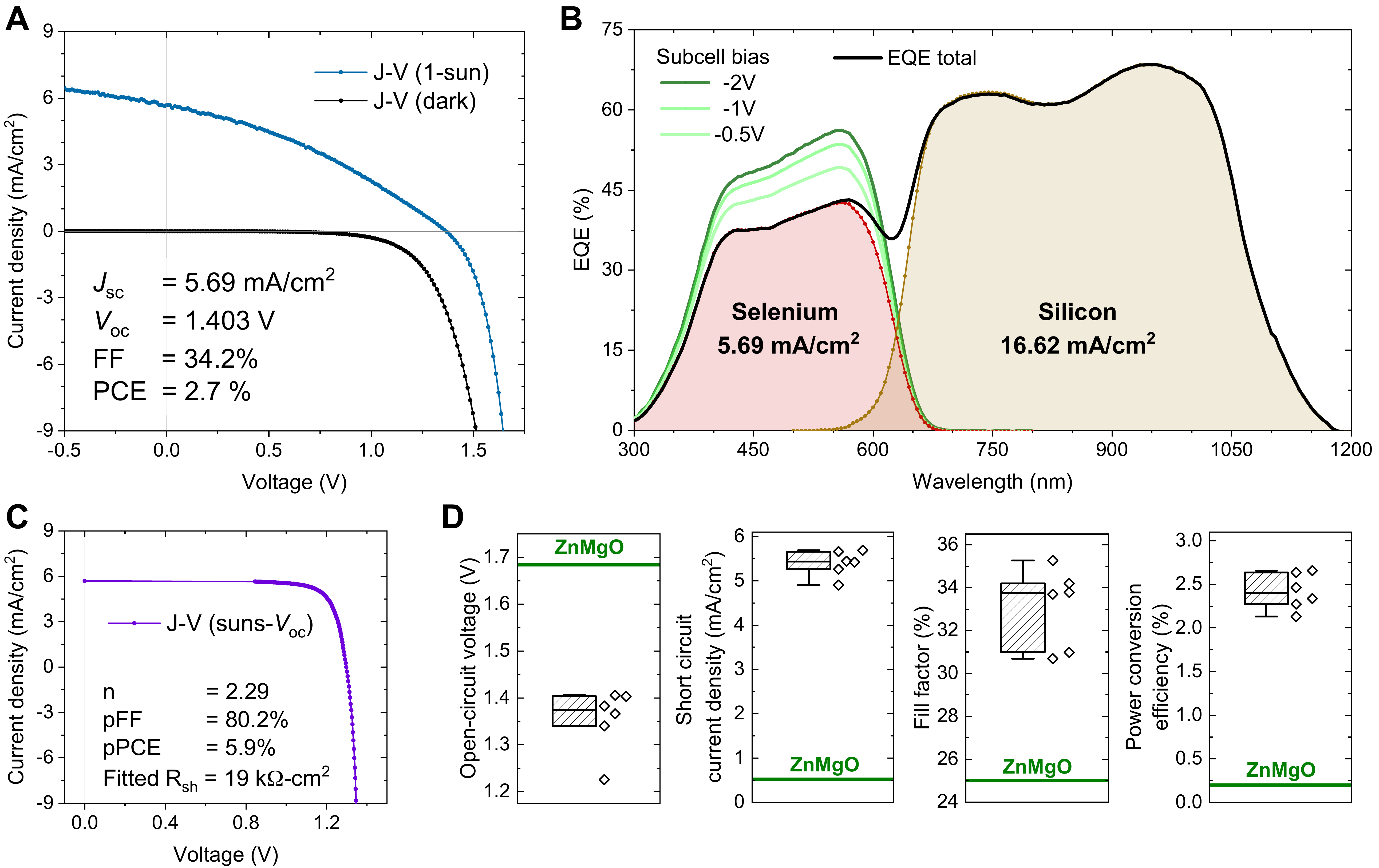}
    \caption{Photovoltaic device performance of the TiO$_\text{2}$-based selenium/silicon tandem solar cell. (\textbf{A}) Current-voltage (\textit{J-V}) curves measured under 1Sun and dark conditions. (\textbf{B}) External quantum efficiency (EQE) spectra of each subcell measured under short-circuit conditions, along with the spectra of the selenium top cell under reverse bias conditions. The AM1.5G-equivalent short-circuit current density of each subcell has been integrated and highlighted in bold. (\textbf{C}) \textit{J-V} curve reconstructed from suns-\textit{V}$_\text{oc}$ measurements, where the effects of series resistance have been eliminated. (\textbf{D}) Batch statistics of the performance metrics of the TiO$_\text{2}$-based tandem devices compared to the ZnMgO-based tandem device.}
    \label{fig:Figure4}
\end{figure*}

To investigate the sensitivity towards the choice of TCO, we synthesize ZnMgO, TiO$_\text{2}$ and poly-Se thin-films and examine their relative energy band positions. The valence band maximum (VBM) and electronic work function (WF) of each sample surface were measured using ultraviolet photoemission spectroscopy, as shown in Fig. \ref{fig:Figure3}A. To minimize the influence of the substrate, which could deplete carriers and significantly shift the Fermi-level in materials with low carrier concentrations\cite{hoye2016a}, the poly-Se thin-film was grown on ZnMgO with a threefold increase in thickness ($\sim$800 nm). To accurately determine the position of the Fermi-level ($E_\text{F}$), a gold leaf was mounted in electrical contact with the edge of each thin-film. Additionally, a small bias of $\text{-5}$V is applied to the sample surface to deconvolute the true work function of the surface from the internal work function of the spectrometer. The width of the spectrum is measured as the difference between the low-energy secondary electron cutoff and the Fermi-level, and the work function is calculated by subtracting the spectrum width from the incident photon energy. The effects of surface photovoltage and Fermi-level pinning have not been considered. To determine the electron affinity of each material, the bandgap must be subtracted from the sum of the work function and valence band offset. The optical bandgaps are extrapolated from Tauc-plots of UV-vis transmission measurements, as shown in Fig. \ref{fig:Figure3}B. It is important to note that the optical characterization is prone to its own set of measurement and analysis errors, e.g. diffuse light scattering, excitonic effects and sub-bandgap defect states.

Using the electron affinity and optical bandgap of each layer, we construct the energy band diagrams of the two heterostructures shown in Fig. \ref{fig:Figure3}C. The structures feature either ZnMgO or TiO$_\text{2}$ as the n-type contact, and an arbitrary TCO with a work function of WF$=$4.4 eV. Ideally, the conduction band offset (CBO) in the pn-heterojunction of a solar cell falls within the range of 0 eV and +0.4 eV, resulting in a small "spike-like" offset\cite{minemoto2001a, gloeckler2005a}. The electron affinity of ZnMgO may be tuned by varying the concentration of Mg, and we find that the elemental composition Mg/(Zn+Mg)$=$0.2 used in this work yields an electron affinity of EA$_\text{ZnMgO}=$3.7 eV, consistent with reported values in literature\cite{takahashi2022a, chantana2023a}. Consequently, this leads to an ideal offset of CBO$=$+0.2 eV at the poly-Se/ZnMgO-interface, which is also indicated by a high built-in potential of V$_\text{bi}=$1.5 V according to capacitance-voltage measurement shown in Fig. S.9. However, the work function difference at the ZnMgO/TCO-interface forms a transport barrier greater than +0.6 eV for the electrons, and the high concentration of Mg is known to significantly increase the resistivity of the thin-film, thereby contributing to parasitic series resistance losses\cite{jamarkattel2022a}. In contrast, TiO$_\text{2}$ possesses a substantially higher electron affinity of EA$_\text{TiO${_2}$}=$4.2 eV, resulting in a "cliff-like" CBO. This cliff-like offset lowers the activation energy of interface recombination\cite{crovetto2017a}, which is expected to significantly reduce the open-circuit voltage in our device. On the other hand, the work function difference at the TiO$_\text{2}$/TCO-interface is less than 0.3 eV, making the electron barrier at this interface benign.

The influence of the transport barrier at the ETL/TCO-interface on the photovoltaic performance is investigated for the two structures in Fig. \ref{fig:Figure3}D and E. Using SCAPS-1D, the illuminated \textit{J-V} curves are simulated while varying the work function of the TCO in the range from 4.1 eV to 4.8 eV. It is important to note that the width of the transport barrier decreases with increasing doping density of the ETL, and a sufficiently high doping may allow for more efficient quantum tunnelling of electrons. In a study by Jamarkattel et al.\cite{jamarkattel2022a}, the n-type conductivity of sputter-deposited ZnMgO was investigated using Hall-effect measurements. As the synthesis conditions and the elemental composition of the ZnMgO thin-films are similar to those used in our study, the doping density for both ZnMgO and TiO$_\text{2}$ is fixed at $N_\text{D}=\text{3}\times\text{10}^\text{18}$ cm$^\text{-3}$, as a conservative estimate. As expected, the ZnMgO-based devices are much more sensitive to the choice of TCO, as the transport barrier is much higher relative to that of the TiO$_\text{2}$-based devices, which are more robust in this range of WF. The work function of FTO is typically reported to be WF$\sim$4.4 eV\cite{andersson1998a, qiao2006a}, whereas the work function of ITO is typically reported to be in range of 4.5-4.8 eV\cite{nehate2018a, park1996a}. Based on these SCAPS-1D simulations, replacing ZnMgO with TiO$_\text{2}$ is expected to reduce the transport barrier height at the ETL/TCO-interface, effectively improving the photovoltaic device performance.

In Fig. \ref{fig:Figure4}A and B, the photovoltaic device performance of the TiO$_\text{2}$-based selenium/silicon tandem device is shown. As expected, the open-circuit voltage is reduced to V$_\text{oc}=$1.40 V, but the fill-factor has improved to FF=34.2\%. This indicates that we have reduced the carrier transport limitations quite significantly. Furthermore, as TiO$_\text{2}$ is sputter-deposited in a reactive atmosphere from a metallic Ti-target with a limited supply of O$_\text{2}$, the thin-film is expected to be oxygen poor. As oxygen vacancies are well-known to act as native n-type dopants in TiO$\text{2}$, and the introduction of Mg to ZnO decreases the carrier concentration, it is expected that the doping density in TiO$\text{2}$ would be significantly higher compared to that of ZnMgO. As a consequence, the depletion region in selenium should be much wider, which is indicated by the dramatic increase in the top cell EQE, and a reduced increase in collection efficiency from applying reverse bias to the top cell. In addition, the transport barrier at the ETL/TCO-interface is expected to be thinner and allow for more efficient electron tunneling as the doping density of the ETL increase. However, while the TiO$_\text{2}$ thin-film may be more conductive, and depletes a wider region of the selenium absorber, the recombination at the poly-Se/TiO$_\text{2}$-interface may be significantly worse. The \textit{J-V} curve reconstructed from suns-\textit{V}$_\text{oc}$ measurements is shown in Fig. \ref{fig:Figure4}C, where the pseudo fill-factor is quantified to be pFF=80.2\%, and the ideality factor is n=2.29, which is close to the best achievable ideality factor for a tandem of n=2. Finally, when comparing the TiO$_\text{2}$-based and the ZnMgO-based tandem devices in Fig. \ref{fig:Figure4}D, both the short-circuit current density and fill factor improved significantly from replacing ZnMgO with TiO$_\text{2}$, resulting in a ten fold increase in the efficiency of the champion device PCE=2.7\%.

We have demonstrated two monolithic selenium/silicon tandem solar cell structures and identified the presence of an electron transport barrier at the ETL/TCO-interface. Such a barrier may restrict the flow of electrons, resulting in a significant reduction in the power conversion efficiency. To address this issue, we replaced ZnMgO as the ETL with TiO$_\text{2}$, which led to a reduction in the barrier height, but compromising the ideal conduction band offset in the pn-heterojunction. While ZnMgO offers a favourable band alignment with poly-Se, its tunability through Mg-alloying is accompanied by increased resistivity. On the other hand, TiO$_\text{2}$ offers better conductivity and a reduced barrier height, but its cliff-like CBO with poly-Se is expected to increase interface recombination losses. Instead of further exploring alloys where bandgap and resistivity changes must be balanced, it may be a more promising approach to incorporate surface modification layers. An ultra-thin molecular or polymeric dipole layer has the capability to alter the contact behaviour with shifts on the order of electron volts\cite{walsh2022a}. This has been studied extensively for use in organic PV, and is currently employed in state-of-the-art perovskite/silicon tandem solar cells\cite{al-ashouri2020a, mariotti2023a}. Alternatively, one could consider replacing ITO with another recombination layer, such as ZnMgO:Al, which may be deposited while maintaining the substrate at room temperature without compromising conductivity\cite{chantana2023a}.

Another issue contributing to the parasitic series resistance is the in-house sputter deposited ITO. When deposited at 300$^\circ$C, the thin-film exhibits a resistivity of 6.4$\times$10$^{-3}$ $\Omega\cdot$cm, but when deposited at 100$^\circ$C, it can be as high as 3.7$\times$10$^{-2}$ $\Omega\cdot$cm according to four point probe measurements. The low melting point of selenium makes it more challenging to synthesize a highly conductive TCO as a front contact, as higher substrate temperatures are typically used to improve the conductivity of the TCO. Even with the applied Au-contact grid to compensate for the expected ohmic losses, the fill factor is still reduced from pFF=80.2\% to FF=34.2\%.

It is however important to underline, that it is a more pressing matter to invert the device polarity as evident from the parallel processed bifacial device. Only a few reports have investigated the inverted device polarity\cite{nielsen2021a, liu2020b}, but instead of fabricating the selenium top cell in a substrate configuration, one could potentially use a direct transfer technique (release/flip/bonding). The low melting point of selenium could be an advantage here, and this solves the issue of synthesizing a conductive TCO on top of selenium. Alternatively, the collection efficiency of the top cell may be improved by increasing the doping density in ZnMgO, thus effectively increasing the depletion width in the absorber, or the diffusion length in the selenium photoabsorber by means of the carrier lifetime or mobility. This point is consistent with our previous publications on improving the optoelectronic quality of the selenium absorber\cite{nielsen2022a, nielsen2023a}. However, prior to undertaking elaborate efforts in defect-engineering of selenium, it is essential to determine the appropriate substrate materials on which to grow the absorber. The results of this work demonstrate the significant impact of using ITO instead of FTO.\\

\section*{Conclusion}

In summary, we have successfully fabricated the first monolithically integrated selenium/silicon tandem solar cells, demonstrating the potential of selenium as a wide bandgap photoabsorber for tandem devices. These devices achieved encouraging results, including an open-circuit voltage of V$_\text{oc}$=1.68 V and a power conversion efficiency of PCE=2.7\%. The analysis of carrier transport losses from suns-\textit{V}$_\text{oc}$ measurements revealed an impressive pseudo fill factor of pFF=80\% and diode ideality factor of n=2.3, emphasizing the importance of reducing parasitic series resistance losses to realize high-efficiency tandem devices. Moreover, using photoemission spectroscopy and SCAPS-1D device simulations, we identified a critical transport barrier at the interface between the ETL and the recombination layer, restricting the flow of charge carriers. By replacing the electron-selective transport material, we achieved a promising 10-fold increase in efficiency, albeit compromising the ideal band alignment at the pn-heterojunction. This highlights the importance of optimizing the heterostructure interfaces, where the inclusion of surface modification layers should be considered.

Furthermore, our study of parallel-processed bifacial selenium single junctions revealed a strong dependence on device polarity, with significantly improved current densities when illuminating through the ETL. This finding is in agreement with a previous report on bifacial selenium solar cells, and emphasize the potential improvements in device performance from inverting the structure and positioning the carrier-separating junction closer to the illuminated surface. To address the still significant open-circuit voltage deficit of the selenium top cell, the optoelectronic quality and thus the diffusion length in selenium should be improved. Moving forward, advancements in carrier-selective contact materials and interfaces, device polarity inversion, reduction of parasitic series resistance, and enhancing the diffusion length in selenium will all be crucial steps towards unlocking the full the potential of selenium/silicon tandem solar cells.\\

\section*{Conflicts of interest}
There are no conflicts of interest to declare.\\

\section*{Acknowledgements}
This work was supported by the Independent Research Fund Denmark (DFF) grant 0217-00333B, and the Villum Foundation V-SUSTAIN grant 9455.\\



\balance


\bibliography{references} 
\bibliographystyle{rsc} 

\end{document}


\maketitle

\paragraph*{Materials} Silicon wafers (Czochralski-grown, n-type/phosphorus, $<$100$>\pm$0.5$^\circ$, 1-20 $\Omega$-cm, 350$\pm$15 $\mu$m, DSP) were purchased from Siegert Wafer. Amorphous selenium (99.999+\%, metals basis) and tellurium (99.9999\%, metals basis) shots were purchased from Alfa Aesar. Ti (99.995\%), Zn$_\text{0.85}$Mg$_\text{0.15}$O (99.95+\%), MoO$_\text{3}$ (99.9\%), ITO 90/10 wt\% (99.99\%), Ag (99.99\%) and Au (99.99\%) sputtering targets were purchased from AJA International. FTO-coated (2.2 mm thick, $\sim7\,\Omega/\text{sq}$) and ITO-coated (1 mm thick, $10-15\,\Omega$) glass substrates were purchased from Sigma-Aldrich and SPI Supplies, respectively.

\paragraph*{Device characterization} Current-voltage (\textit{J-V}) measurements of photovoltaic devices are measured using a Keithley 2561A source meter with 4-terminal sensing under 1 sun illumination (Newport 94082A solar simulator, class ABA, equipped with a 1600 W Xe arc lamp and appropriate AM1.5G filters). The light intensity is calibrated in the plane of the device under test using a reference solar cell from Orion. As no mask aperture is used during the acquisition, the active area is determined by calculating the AM1.5G equivalent current density using the integral of the external quantum efficiency (EQE) spectrum of the device under test. EQE spectra are measured using the QEXL from PV Measurements calibrating using a silicon reference photodiode. Photoconductance decay measurements are aquired using the WCT-120 Photoconductance Lifetime Tester from Sinton Instruments. The instrument uses a xenon flash lamp with optical filters and an eddy-current conductance sensor to measure the carrier lifetime. Suns-\textit{V}$_\text{oc}$ measurements are conducted using the accessory stage to the WCT-120 from Sinton Instruments. The setup features an illumination sensor calibrated from 0.006 to 6 suns, and the sample chuck is temperature controlled (25$^\circ$C).

\paragraph*{Additional characterization} Scanning electron microscopy (SEM) images are acquired using a Supra 40 VPSEM from Zeiss. UV-vis transmission spectra are measured at room temperature using a UV-2600 spectrophotometer from Shimadzu. Ultraviolet photoemission spectroscopy measurements are carried out in the commercial Thermofisher Scientific Nexsa XPS system, using a helium discharge lamp as the photon source with principal photon energies 21.2 eV (He I) and 40.8 eV (He II). Variable angle spectroscopic ellipsometry measurements are carried out using a M2000XI-210 ellipsometer from J.A. Woollam Co., Inc.

\begin{figure*}[b!]
    \centering
    \begin{subfigure}[b]{0.3\textwidth}
        \centering
        \includegraphics[width=0.8333333\textwidth,trim={352 160 359 164},clip]{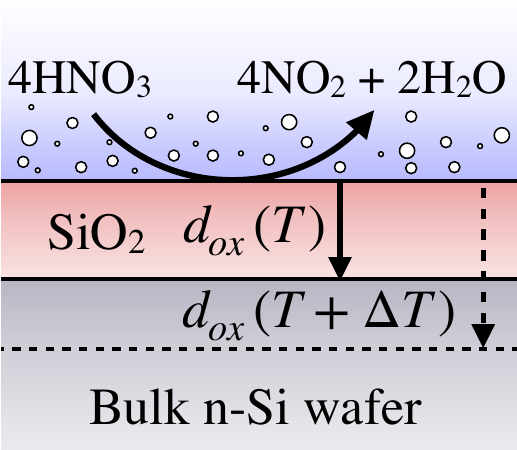}
        \caption{Nitric acid oxidation of silicon.}
        \label{fig:HNO3_1}
    \end{subfigure}
    \hspace{0.15cm}
    \begin{subfigure}[b]{0.3\textwidth}
        \centering
        \includegraphics[width=0.8333333\textwidth,trim={352 160 359 164},clip]{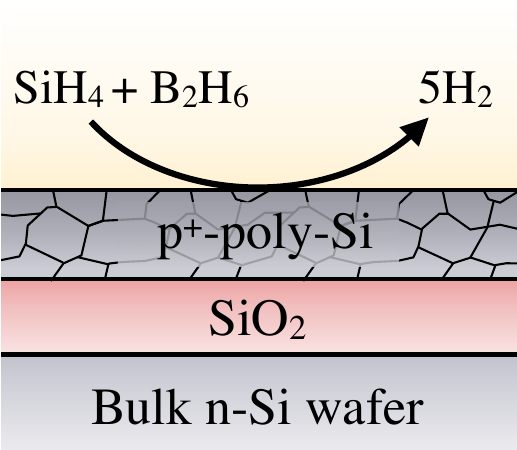}
        \caption{LPCVD deposition of poly-Si.}
        \label{fig:HNO3_2}
    \end{subfigure}
    \hspace{0.15cm}
    \begin{subfigure}[b]{0.3\textwidth}
        \centering
        \includegraphics[width=0.8333333\textwidth,trim={352 160 359 164},clip]{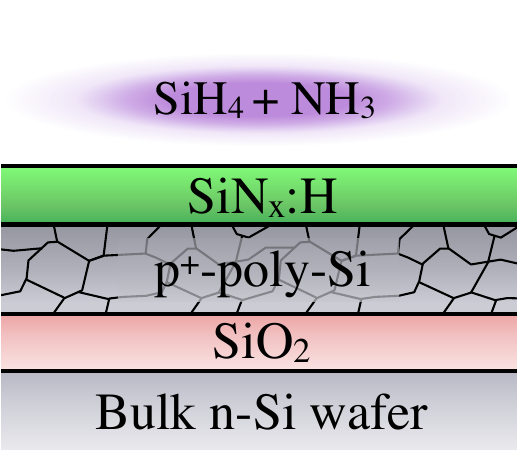}
        \caption{PECVD deposition of SiN$_\text{x}$.}
        \label{fig:HNO3_3}
    \end{subfigure}    \caption[Thin-film deposition techniques used to fabricate silicon bottom cell precursors]{Schematics of the thin-film deposition techniques involved in the processing of the p$^{\text{+}}$n junction of the silicon bottom cell precursors. The same techniques are used to form the passivated n$^{\text{+}}$n junction on the rear side of the wafer, with the exception of phosphine (PH$_\text{3}$) being used as the dopant gas during the LPCVD of poly-Si.}
    \label{fig:HNO3_1_and_2}
\end{figure*}

\begin{figure*}[b!]
    \centering
    \includegraphics[width=1\textwidth,trim={0 0 0 0},clip]{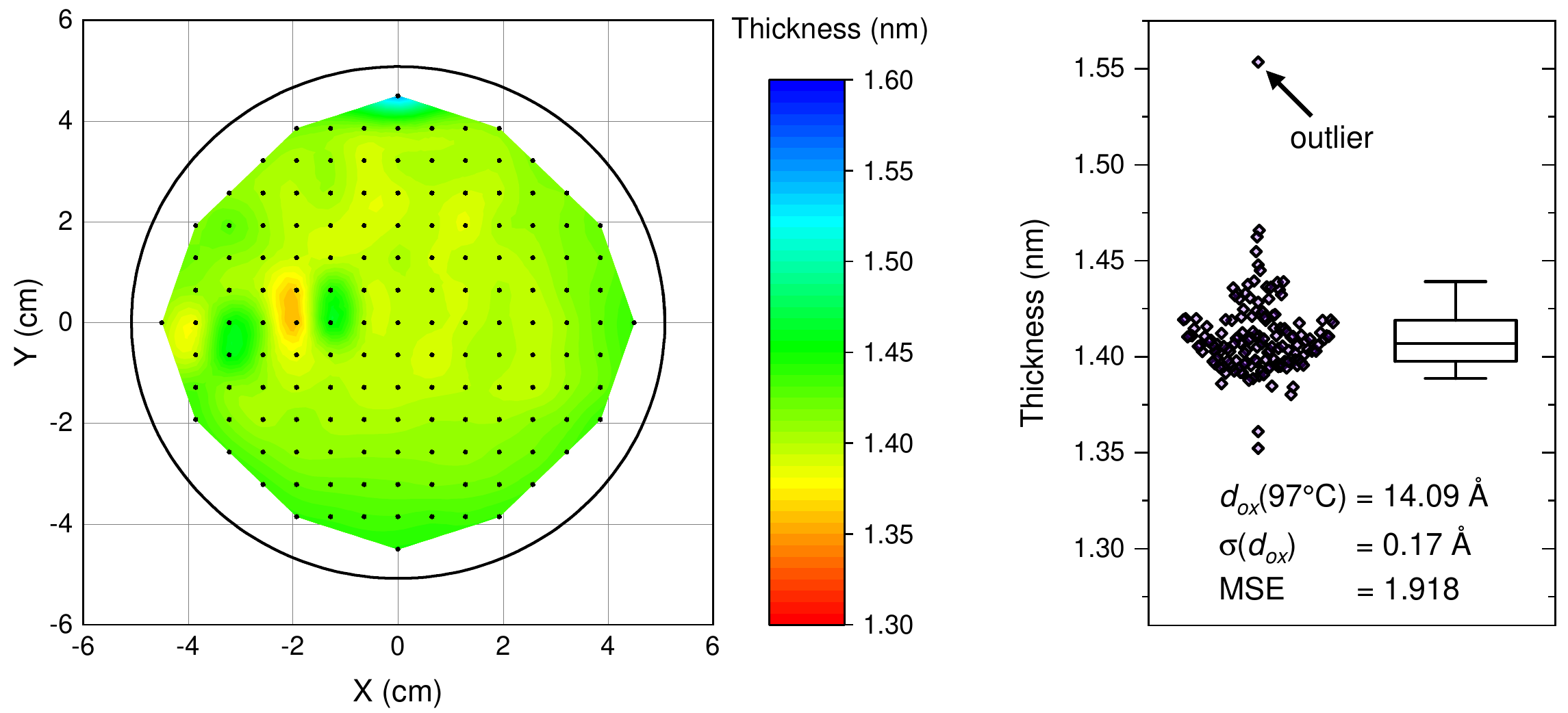}
    \caption{Spectroscopic ellipsometry measurements mapping out the thickness of the tunnel oxide grown in hot nitric acid on a 4 inch n-Si wafer. Each data point corresponds to spectral fitting in the region 1.3 eV to 5.9 eV at angles of incidence 55$^\circ$, 60$^\circ$, and 65$^\circ$. The optical constants of the silicon substrate and the Si/SiO$_\text{2}$ interface are retrieved from Hertzinger et al.\cite{Herzinger}}
    \label{fig:NaosThicknessMap}
\end{figure*}

\clearpage

\begin{figure*}[b!]
    \centering
    \includegraphics[width=\textwidth,trim={0 0 0 0},clip]{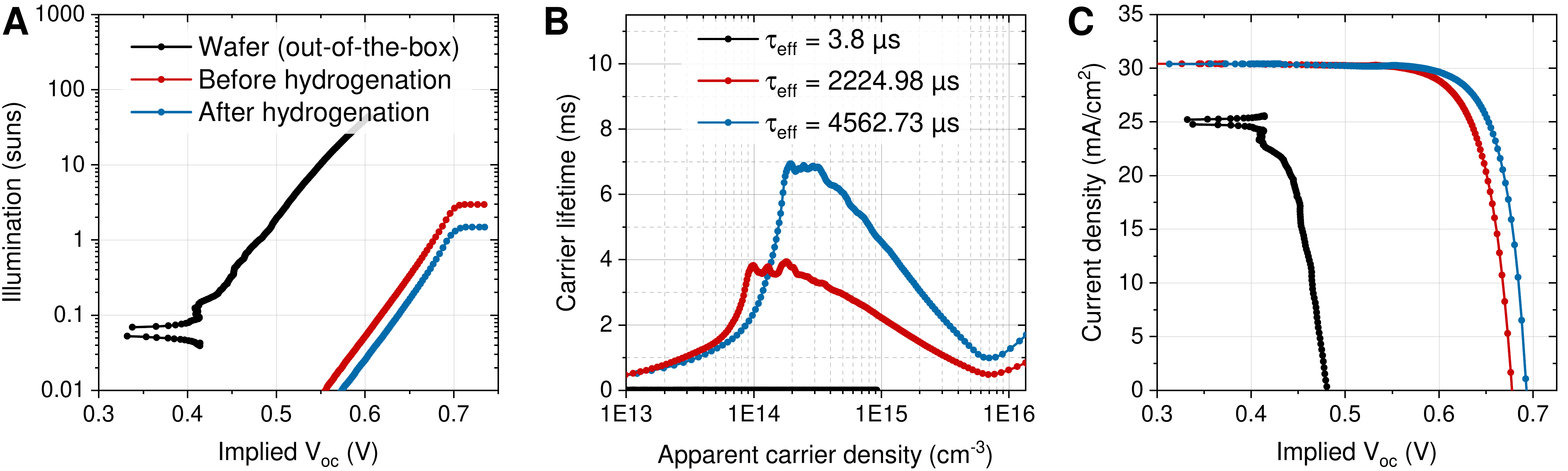}\\
    \caption{Investigating the effect of hydrogenation using injection-level-dependent open-circuit voltage measurements on a device with symmetrical polarity poly-Si(p$^{++}$ )/c-Si(n)/poly-Si(p$^{++}$).}
    \label{fig:ESI3}
\end{figure*}

\begin{figure*}[b!]
    \centering
    \includegraphics[width=\textwidth,trim={0 0 0 0},clip]{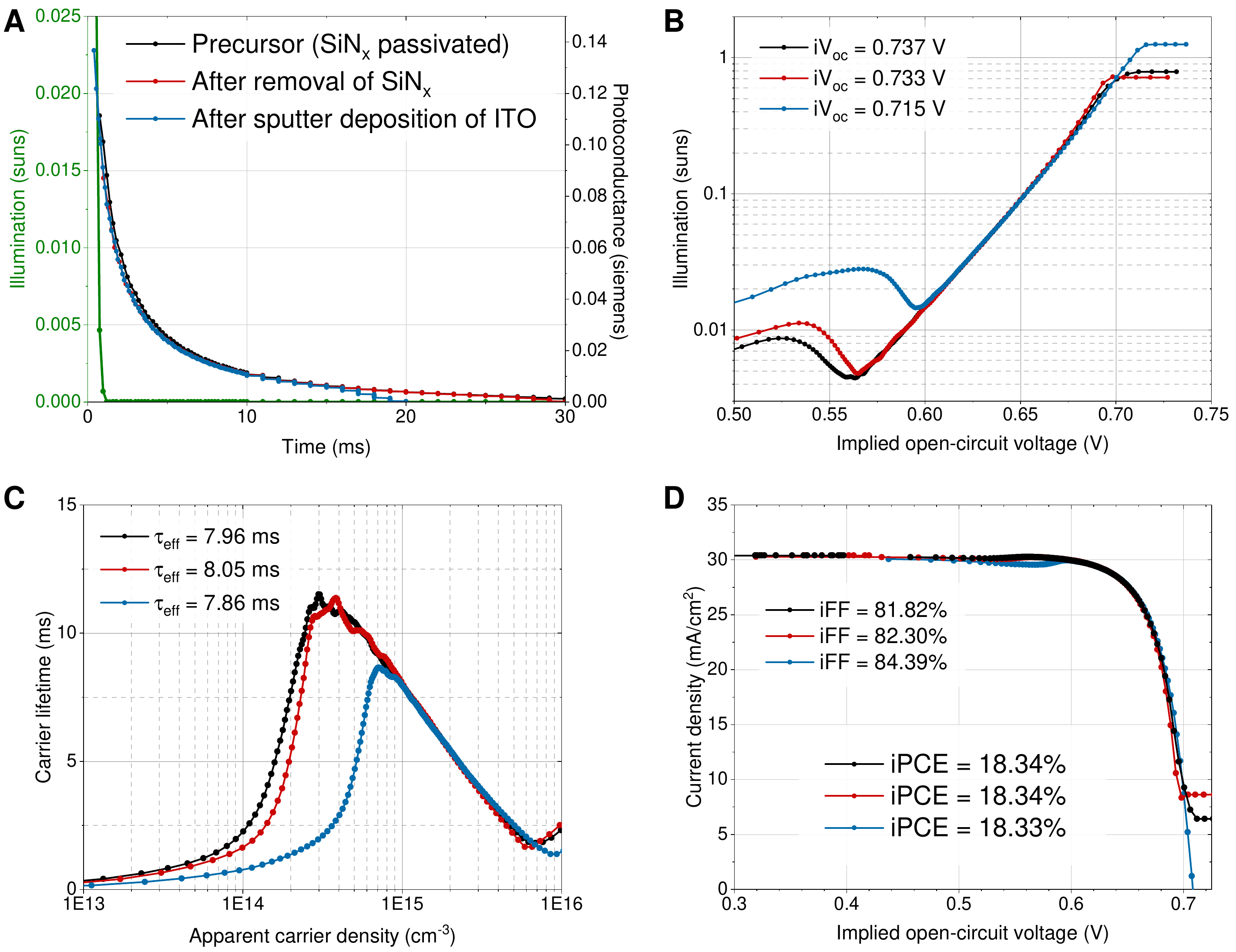}
    \caption{Investigating the effect of sputtering ITO on a silicon bottom cell wafer using photoconductance decay measurements. As evident from the carrier lifetimes and implied \textit{J-V} curves, the sputter damage on the bottom cells is negligible.}
    \label{fig:ESI4}
\end{figure*}

\clearpage

\begin{figure*}[b!]
    \centering
    \includegraphics[width=\textwidth,trim={0 0 0 0},clip]{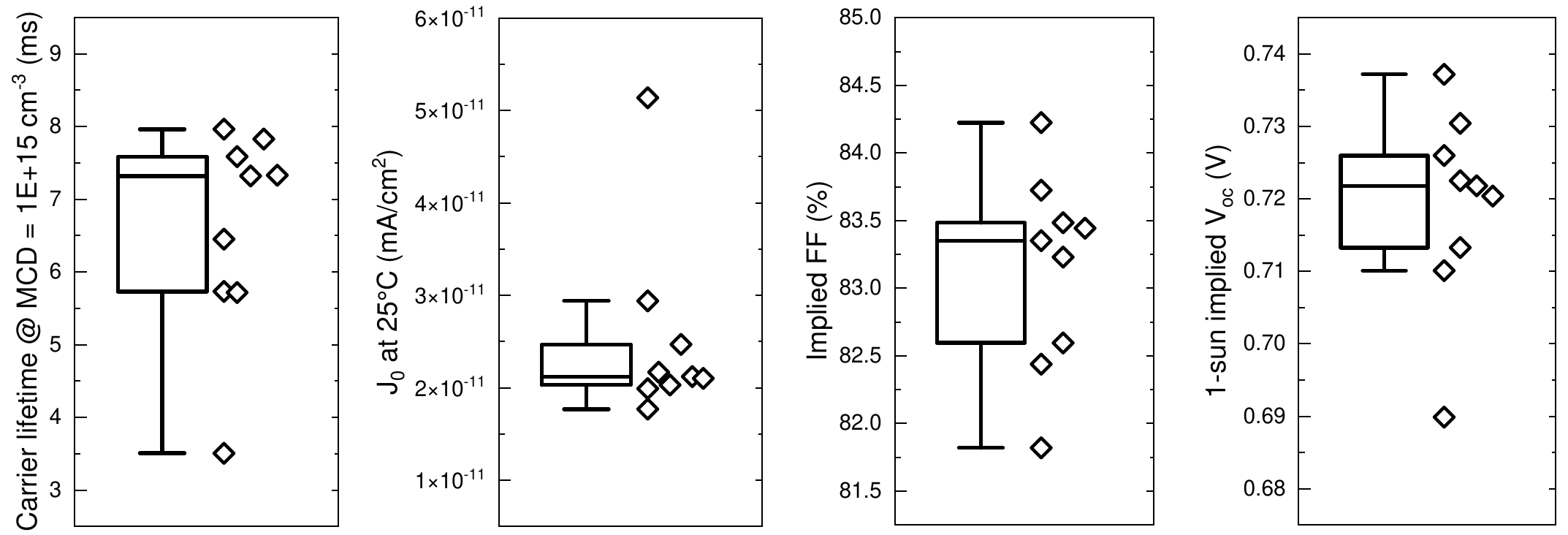}
    \caption{Batch statistics of performance metrics retrieved from photoconductance decay measurements on the silicon bottom cell wafers before the sputter deposition of ITO and the subsequent dicing.}
    \label{fig:ESI2}
\end{figure*}

\begin{figure*}[b!]
    \centering
    \includegraphics[width=\textwidth,trim={0 2551 0 0},clip]{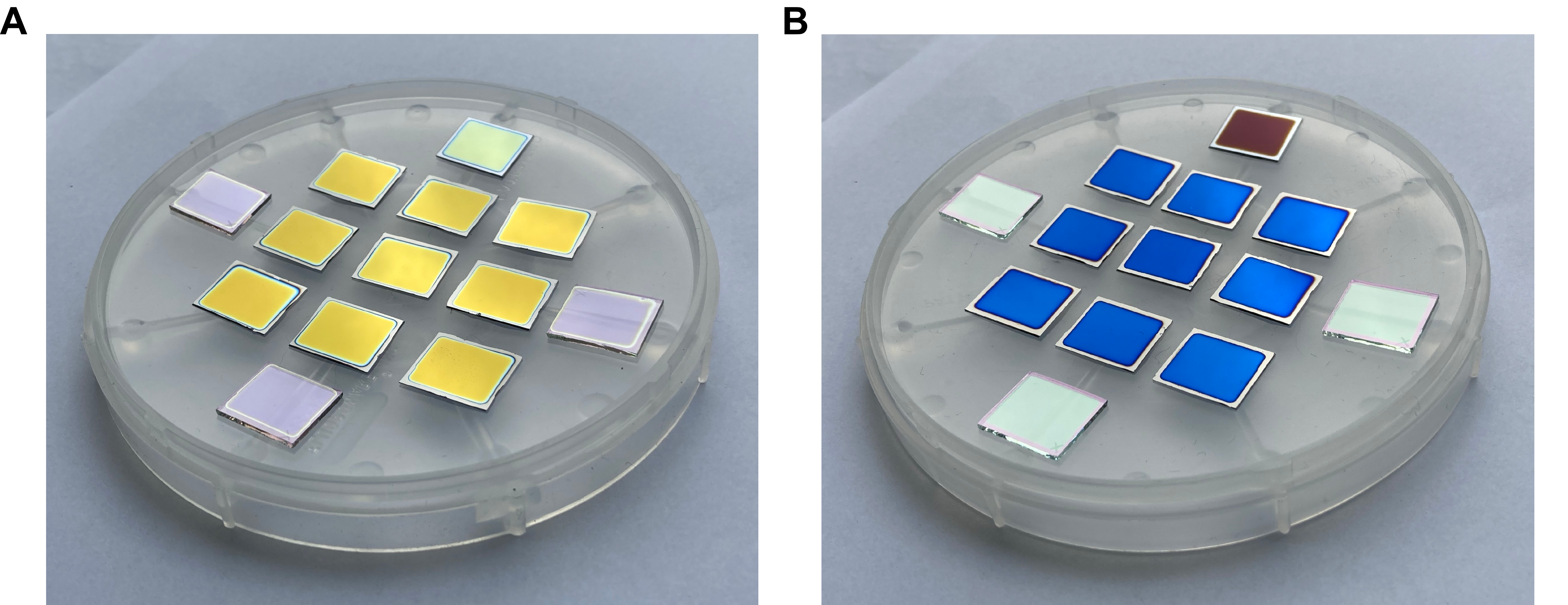}
    \caption{Optical photographs of the two batches of devices after depositing either the oxygen-poor TiO$_\text{2}$ (left) or ZnMgO (right). The outermost semitransparent samples are ITO/glass substrates used for parallel-processed single junction bifacial devices. The top-most sample is a bare silicon substrates used for ellipsometry measurements.}
    \label{fig:ESI6}
\end{figure*}

\clearpage

\begin{figure*}[b!]
    \centering
    \includegraphics[width=\textwidth,trim={0 2602 0 0},clip]{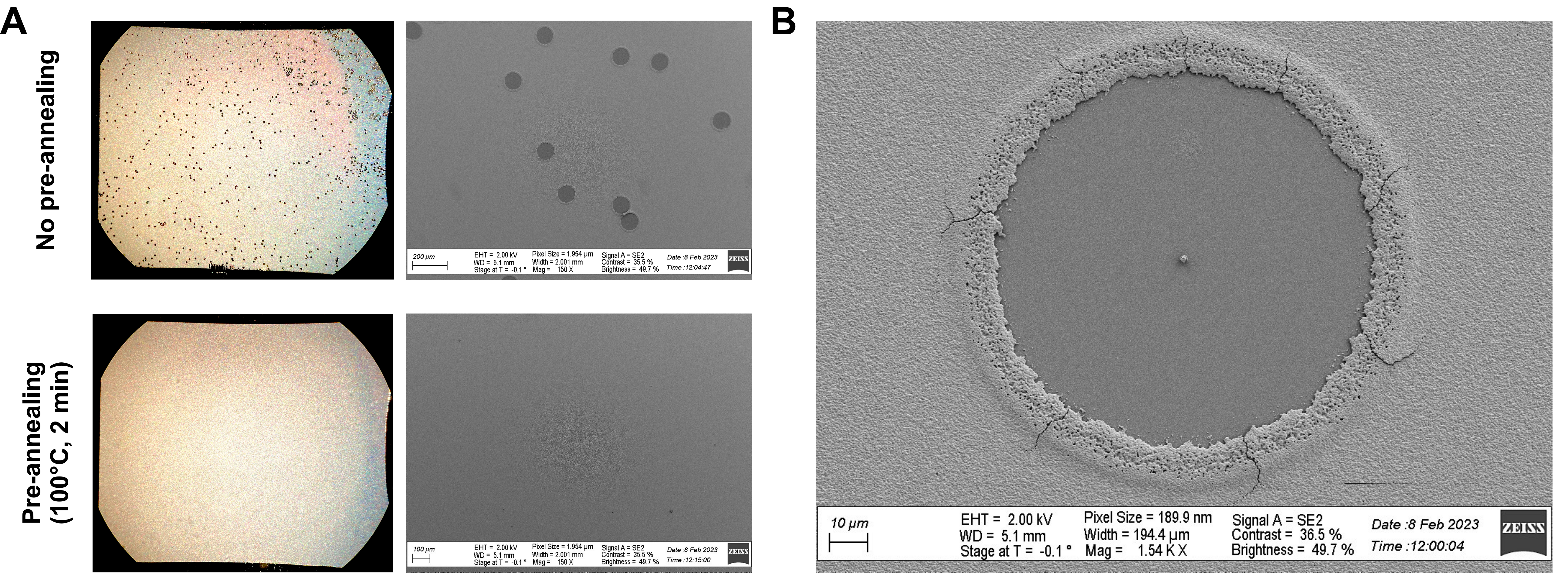}
    \caption{Optical microscopy and SEM-images of the crater-like holes formed in selenium on ZnMgO-based devices on silicon. An additional pre-annealing step at 100$^\circ$C for 2 min was introduced to avoid the formation of such pinholes.}
    \label{fig:ESI5}
\end{figure*}

\begin{figure*}[b!]
    \centering
    \includegraphics[width=0.48\textwidth,trim={0 1884 0 0},clip]{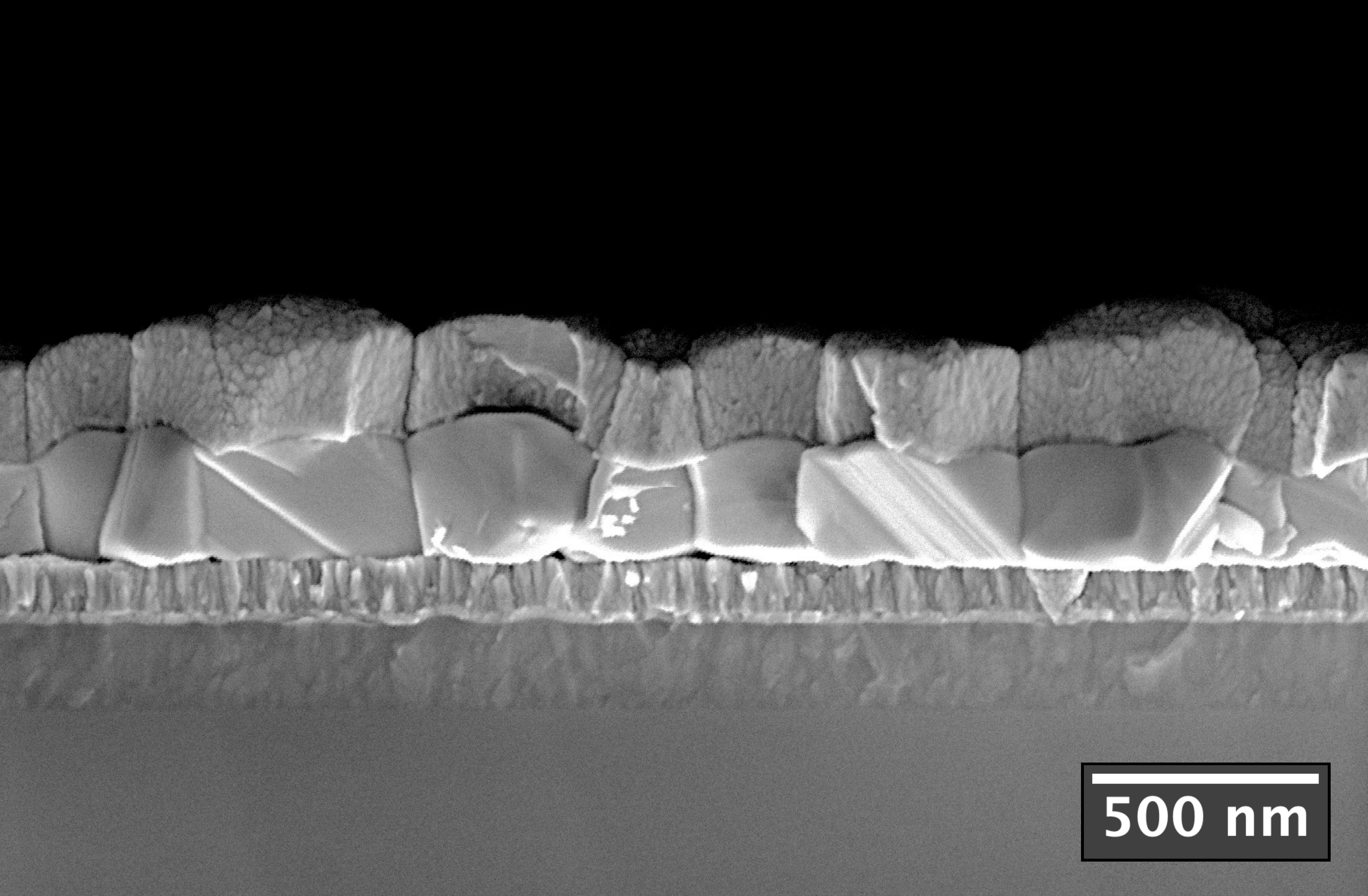}
    \hspace{0.2cm}
    \includegraphics[width=0.48\textwidth,trim={0 1884 0 0},clip]{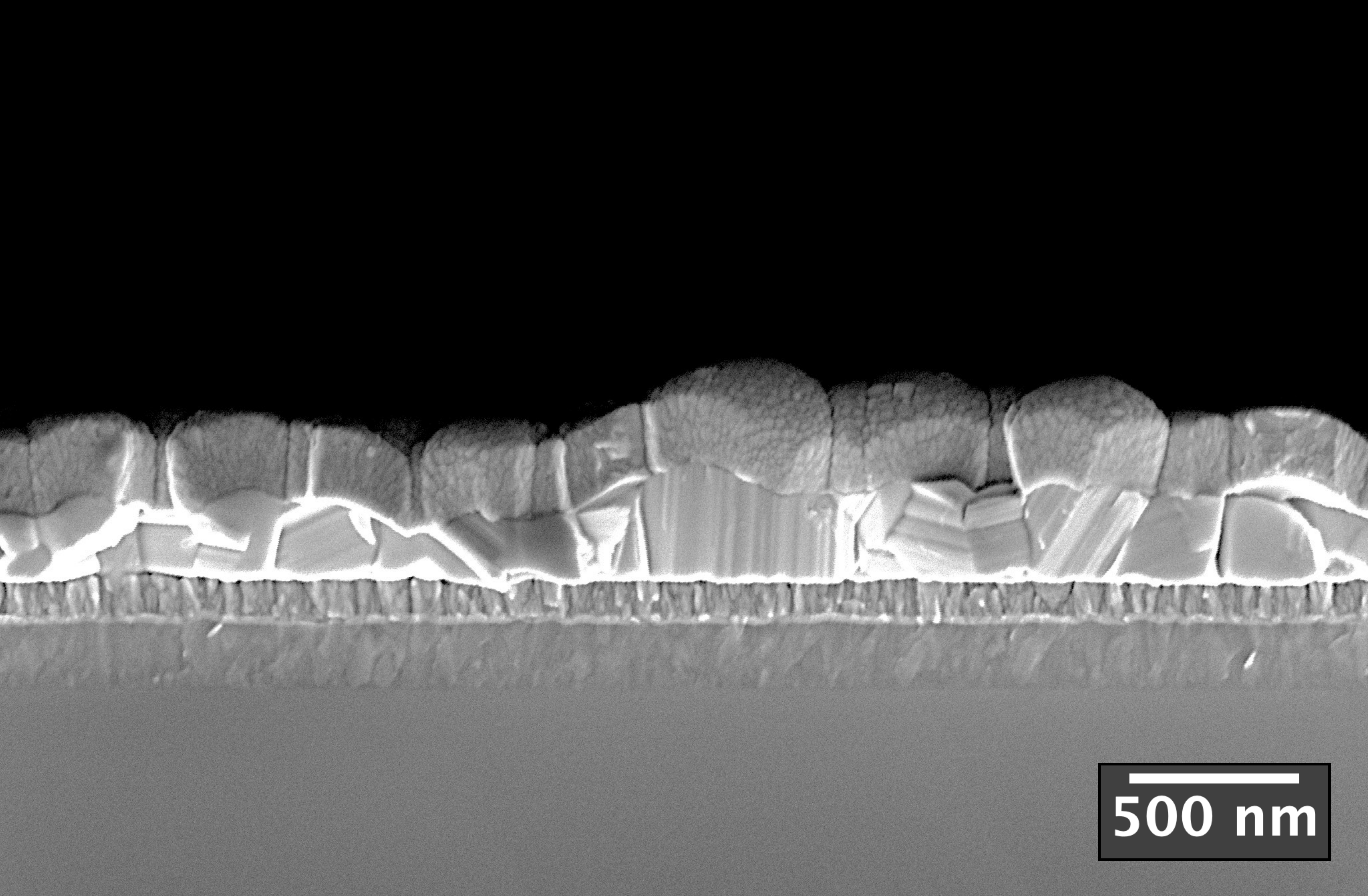}
    \vspace{0.2cm}
    \includegraphics[width=0.48\textwidth,trim={0 1884 0 0},clip]{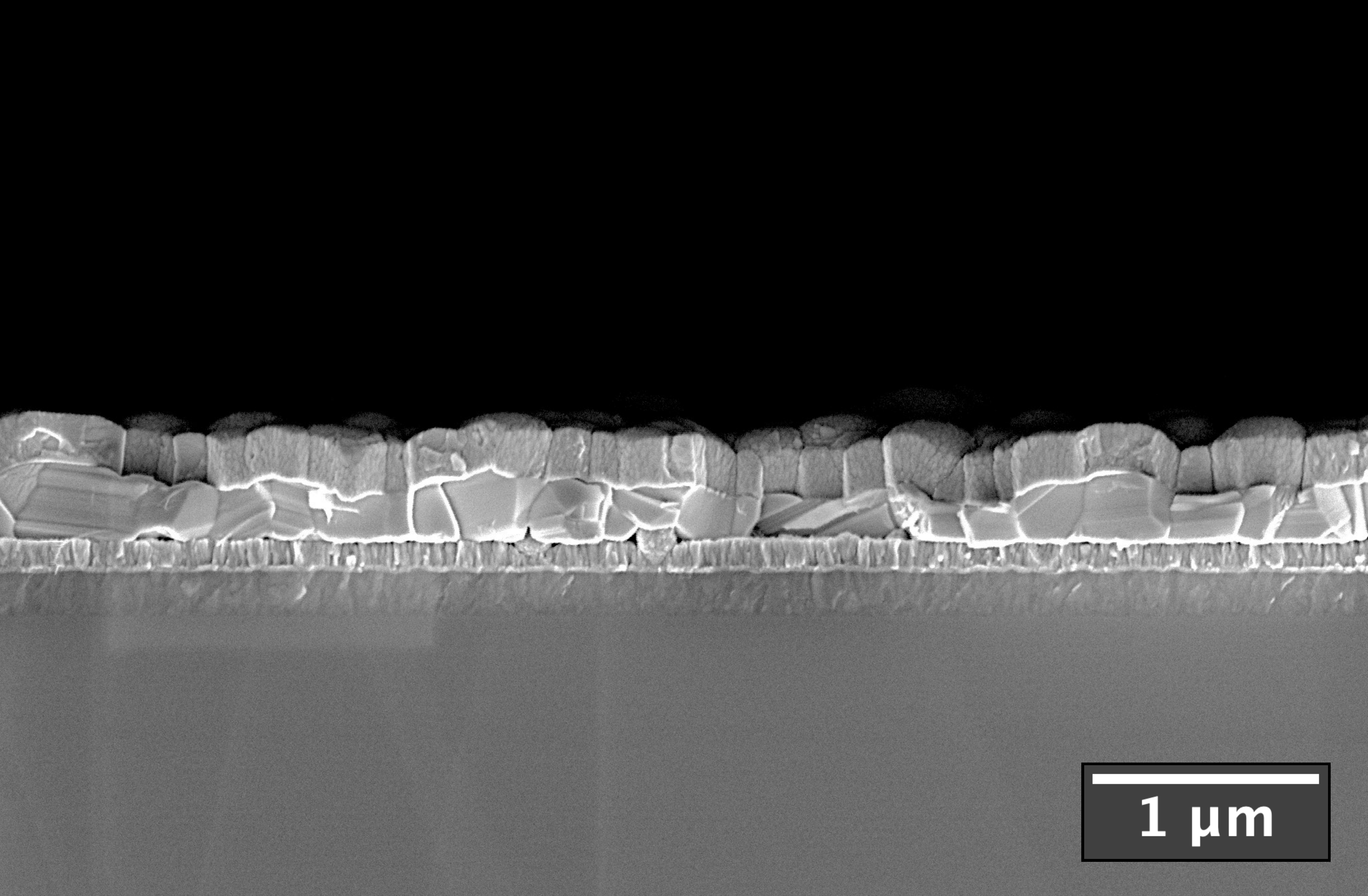}
    \hspace{0.2cm}
    \includegraphics[width=0.48\textwidth,trim={0 1884 0 0},clip]{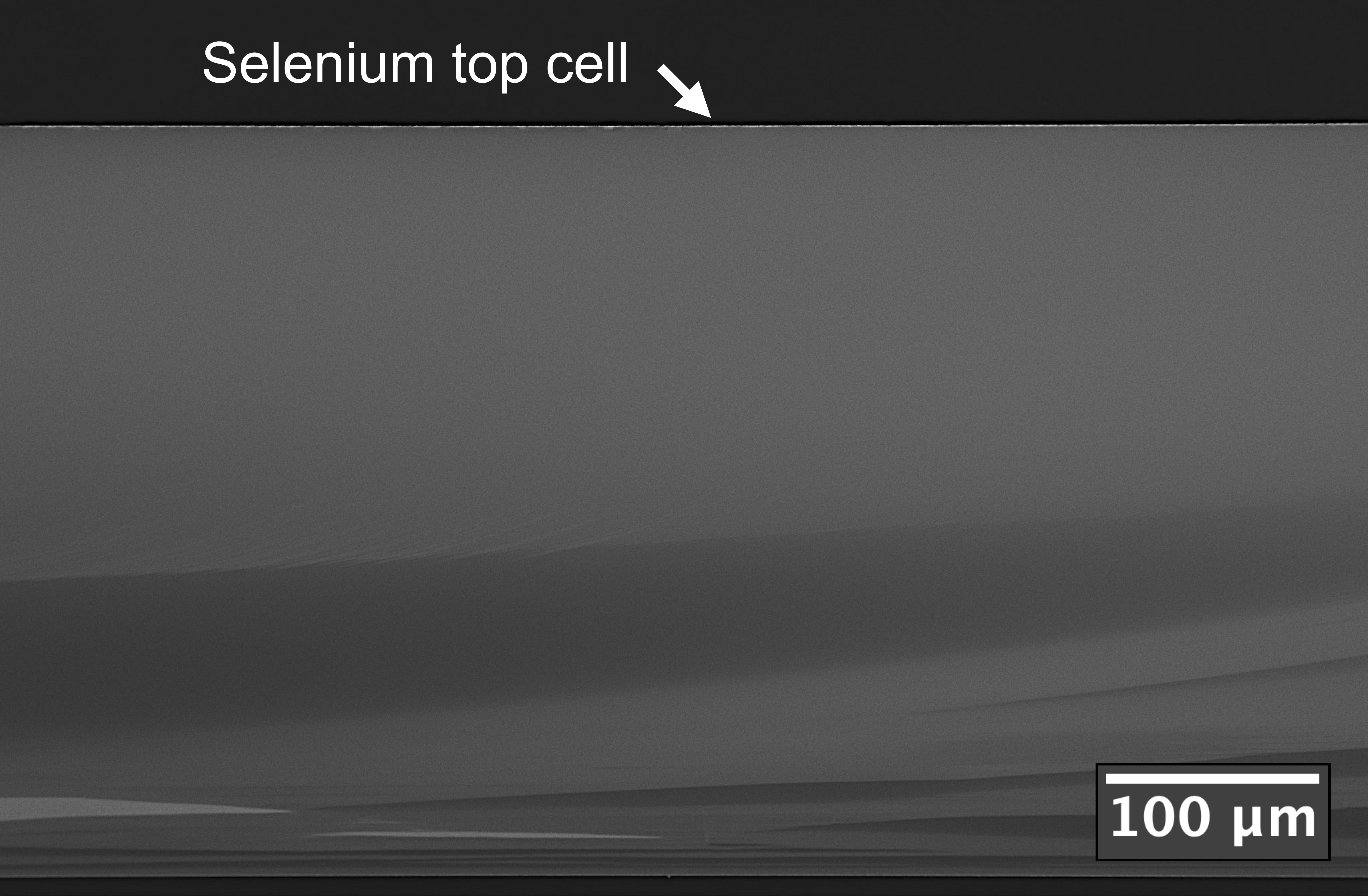}
    \caption{Additional cross-sectional SEM-images of the monolithic selenium/silicon tandem device at various magnifications.}
    \label{fig:ESI1}
\end{figure*}

\clearpage

\begin{figure*}[b!]
    \centering
    \includegraphics[width=0.7\textwidth,trim={0 0 0 0},clip]{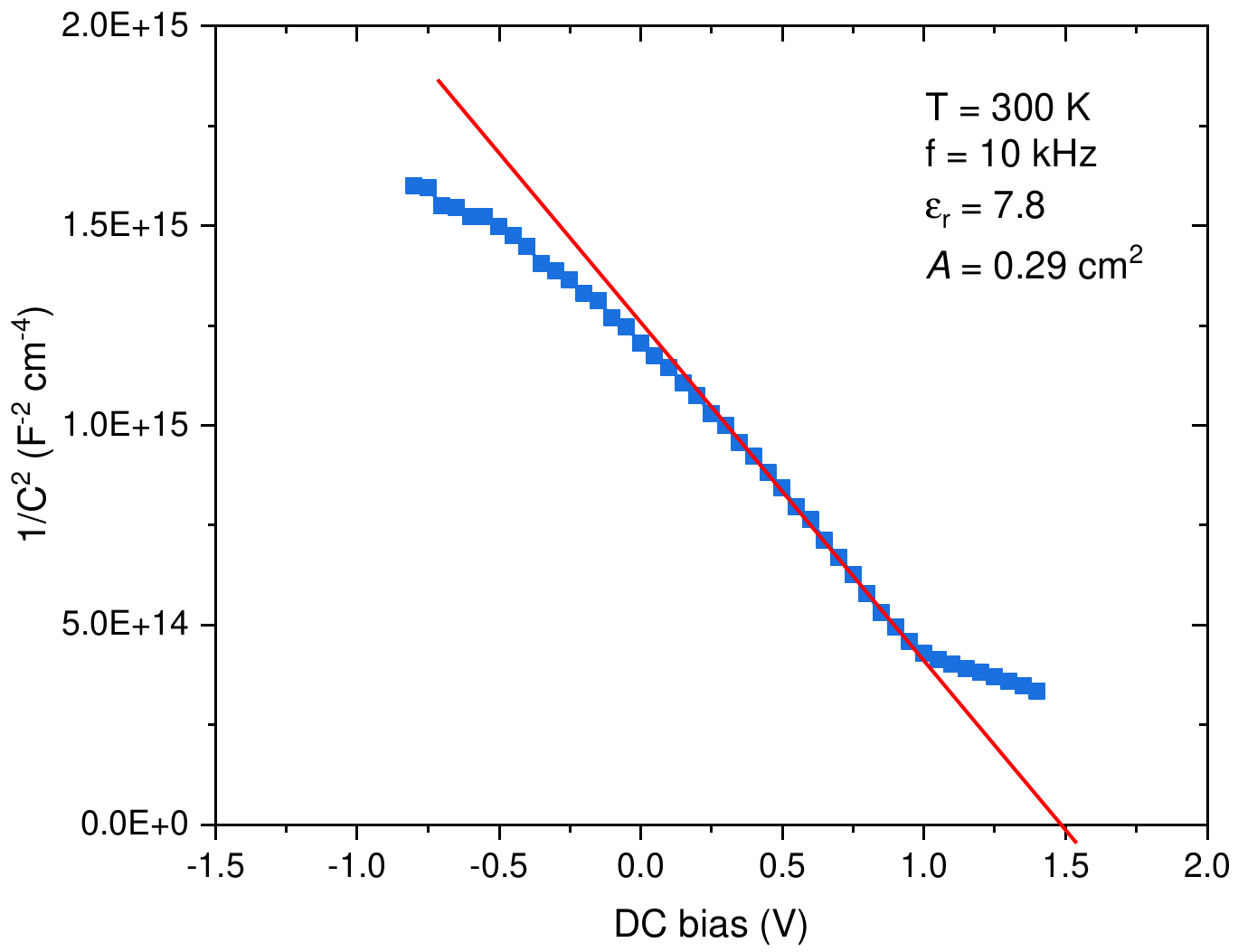}
    \caption{Mott-Schottky plot from C-V measurements of a parallel-processed single junction bifacial device using ZnMgO as the n-type contact. The built-in voltage V$_\text{bi}$=1.5 V is determined from the intercept with the \textit{x}-axis.}
    \label{fig:NaosThicknessMap}
\end{figure*}

\clearpage

\begin{table}[ht!]
\small
  \caption{\ Baseline parameters used in SCAPS-1D device simulations.}
  \label{tbl:example}
  \begin{tabular*}{\textwidth}{@{\extracolsep{\fill}}lrrr}
    \hline \vspace{-0.25cm}\\
    \textbf{Contact properties} & \textbf{Front} &  & \textbf{Back} \\
    $S_\mathrm{e}$ (cm/s) & $\text{10}^{\text{7}}$ & & $\text{10}^{\text{5}}$ \\
    $S_\mathrm{h}$ (cm/s) & $\text{10}^{\text{5}}$ & & $\text{10}^{\text{7}}$ \\
    $\phi$ (eV)\vspace{0.2cm} & $\text{4.4}$ (varied) & & $\text{5.6054}$ (flat bands) \\
    \hline \vspace{-0.25cm}\\
    \textbf{Layer properties} & \textbf{ZnMgO}\cite{Skink, Hawaii} & \textcolor{blue}{\textbf{TiO$_\text{2}$}$^\ddag$} & \textbf{poly-Se}\cite{Origin} \\
    Thickness (nm) & $\text{65}$ & $\text{65}$ & $\text{300}$ \\
    $E_\mathrm{g}$ (eV) & $\text{3.55}$ & \textcolor{blue}{$\text{3.40}$} & $\text{1.95}$ \\
    $\chi$ (eV) & 3.69 & \textcolor{blue}{4.40} & 3.89 \\
    $\epsilon_\mathrm{r}$ & 8.5 & 8.5 & 7.8 \\
    $N_C$ (cm$^{-3}$) & $\text{2.96}\times \text{10}^{\text{18}}$ & $\text{2.96}\times \text{10}^{\text{18}}$ & $\text{8.70}\times \text{10}^{\text{18}}$ \\
    $N_V$ (cm$^{-3}$) & $\text{3.24}\times \text{10}^{\text{18}}$ & $\text{3.24}\times \text{10}^{\text{18}}$ & $\text{1.64}\times \text{10}^{\text{20}}$ \\
    $\mu_\mathrm{e}$ (cm$^{2}$/Vs) & 2.4 & 2.4 & 5$^*$ \\
    $\mu_\mathrm{h}$ (cm$^{2}$/Vs) & 2.4$^\dag$ & 2.4 & 5$^*$ \\
    $N_\mathrm{d/a}$ (cm$^{-3}$)\vspace{0.2cm} & $N_\mathrm{d}= \text{3.2} \times \text{10}^{\text{18}}$ & $N_\mathrm{d}= \text{3.2} \times \text{10}^{\text{18}}$ & $N_\mathrm{a}= \text{1.3} \times \text{10}^{\text{16}}$ \\
    \hline \vspace{-0.25cm}\\
    \textbf{Defect states} & \textbf{ZnMgO} & \textbf{TiO$_\text{2}$} & \textbf{poly-Se}$^\mathsection$ \\
    Type & - & - & Neutral \\
    Energy distribution & - & - & Single level \\
    $N_\mathrm{t}$ (cm$^{-3}$) & - & - & $\text{4.6} \times \text{10}^{\text{16}}$ \\
    $E_\mathrm{t}$ (eV) & - & - & $\textit{E}_\text{V} + \text{0.6}$ \\
    $\sigma_\mathrm{e}$ (cm$^{2}$) & - & - & $\text{10}^{-\text{15}}$ \\
    $\sigma_\mathrm{h}$ (cm$^{2}$)\vspace{0.2cm} & - & - & $\text{10}^{-\text{15}}$ \\
    \hline \vspace{-0.25cm}\\
    \textbf{Interface defect} & \textbf{ZnMgO/Se} & \textbf{TiO$_\text{2}$/poly-Se} & \\
    Type & - & - & \\
    Energy distribution & - & - & \\
    $E_\mathrm{t}$ (eV) & - & - & \\
    $N_\mathrm{t}$ (cm$^{-3}$) & - & - & \\
    $\sigma_\mathrm{e}$, $\sigma_\mathrm{h}$ (cm$^{2}$) & - & - & \\
    \qquad \vspace{-0.25cm} \\
    \hline \vspace{-0.25cm}\\
  \end{tabular*}
  \caption*{\small $^*$ \footnotesize should be considered an upper limit, as the mobility sum $\mu_\Sigma$ is used as the value for both $\mu_\mathrm{e}$ and $\mu_\mathrm{h}$ \cite{Origin}. \\ \small $^\dag$ \footnotesize minority carrier mobility assumed to be similar to the majority carrier mobility from Hall-effect measurements\cite{Hawaii}. \\ \small \textcolor{blue}{$^\ddag$} \footnotesize the only difference between ZnMgO and TiO$_\text{2}$ accounted for here is the optical bandgap and electron affinity to isolate the effect of the interfacial transport barriers. The two parameters have been \textcolor{blue}{highlighted in blue}. \\ \small $^\mathsection$ \footnotesize The neutral defect density in selenium has been adjusted to achieve an effective carrier lifetime $\tau_\text{eff}=$2.2 ns and a diffusion length of $L=$170 nm.}
  \vspace{-0.4cm}
\end{table}

\vfill

\providecommand{\newblock}{}